# Large Teams Overshadow Individual Recognition

**Authors:** Lulin Yang[1], Donna K. Ginther[2,3*], Lingfei Wu[1*]

**Affiliations:**

[1] School of Computing and Information, The University of Pittsburgh, 135 N Bellefield Ave, Pittsburgh, PA 15213

[2] Department of Economics and Institute for Policy & Social Research, University of Kansas, Lawrence, KS, USA

[3] National Bureau of Economic Research, Cambridge, MA, USA

*Corresponding author. E-mail: liw105@pitt.edu (L.W) and dginther@ku.edu (D. G.)

## Abstract

In an ideal world, every scientist's contribution would be fully recognized, driving collective scientific progress. In reality, however, only a few scientists are recognized and remembered. Sociologist Robert Merton first described this disparity between contribution and recognition as the Matthew Effect, where citations disproportionately favor established scientists, even when their contributions are no greater than those of junior peers. Merton's work, however, did not account for coauthored papers, where citations acknowledge teams rather than individual authors. How do teams affect reward systems in science? We hypothesize that teams will divide and obscure intellectual credit, making it even harder to recognize individual contributions. To test this, we developed and analyzed the world's first large-scale observational dataset on author contributions, derived from LaTeX source files of 1.6 million papers authored by 2 million scientists. We also quantified individual credits within teams using a validated algorithm and examined their relationship to contributions, accounting for factors such as team size, career stage, and historical time. Our findings confirm that teams amplify the Matthew Effect and overshadow individual contributions. As scientific research shifts from individual efforts to collaborative teamwork, this study highlights the urgent need for effective credit assignment practices in team-based science.

## Introduction

The progress of science relies on recognizing individual contributions. In an ideal world, every scientist's efforts would be fully acknowledged, motivating further contributions. In reality, however, only a few scientists are recognized and remembered. This gap between ideal and reality reveals that we do not fully understand how the reward system in science works—or how to make it function as intended.

In the 1960s, sociologist Robert Merton examined the relationship between contribution and recognition, using citation impact as a proxy for recognition. He observed that citations disproportionately favor established scientists, even when their contributions are no greater than those of junior peers. He termed this phenomenon the Matthew Effect in science—those already recognized tend to gain even more citations (1). Merton explained this as a result of selective memory: while a preference for citing established scientists may overlook emerging researchers and discoveries, it helps save the collective memory of the scientific community.

In other words, the long-tail, unequal distribution of paper citations is an inevitable outcome of scarce attention and memory.

Merton's work, however, did not address team-based research, where citations recognize papers as collective products rather than an individual's contributions. This was likely for two reasons. First, during Merton's time, most scholarly work was still solo-authored, making the role of teams in the contribution-credit paradox less urgent. However, the landscape of scientific research has dramatically shifted from individual efforts to collaboration (2, 3): solo authorship declined from nearly 80% in the 1960s in sociology, economics, and computer science to 50%, 26%, and 7% in 2010s, respectively (4–6).

Second, there were no methods available in Merton's era to identify individual contributions within teams or how recognition was divided among team members. This limitation has since been addressed through advancements in data and research methodologies. Recent research has developed various ways to calculate credit division within teams (7–12), with the work of Shen and Barabási being one of the most well-validated examples (10). Other studies have utilized self-reported contributions to infer author roles (13–17). In this work, we developed and analyzed the first large-scale observational dataset on author contributions from LaTeX source code, building on the dataset and methodology introduced in our previous research (Pei et al., manuscript in preparation). These advances allow us to overcome past limitations and examine the evolving dynamics of author contribution and recognition in team-based science. Building upon these advances, we ask: How do teams influence reward systems in science?

We hypothesize that while teamwork enables individuals to pool knowledge in productive ways (15), it also divides and obscures credit, making it difficult for outsiders to identify individual contributions within a team (5). This issue could be further exacerbated as team sizes grow, complicating efforts by the scientific community to distinguish the "main brain" behind the work (13) from those in technical or supportive roles (18, 19).

To see these insights in action, consider the most prestigious scientific award, the Nobel Prize. It was established during a time when science was primarily driven by lone geniuses such as Albert Einstein, Marie Curie, and Alfred Nobel himself. However, as science has evolved, major discoveries increasingly require the collaborative efforts of large teams. For example, the CERN collaboration involved 2,932 authors in the paper announcing the discovery of the Higgs boson (20), while the LIGO Collaboration, comprising approximately 1,004 authors, reported the detection of gravitational waves (21). These groundbreaking discoveries were eventually recognized with Nobel Prizes, yet only a few individuals were honored. In 2013, François Englert and Peter Higgs received the Nobel Prize in Physics for the theoretical framework leading to the discovery of the Higgs boson, and in 2017, Rainer Weiss, Barry C. Barish, and Kip S. Thorne were recognized for their decisive roles in the LIGO project. These cases support our hypothesis that properly recognizing all team members involved in large collaborative research becomes increasingly challenging.

Here, we address the challenge of empirically examining the relationship between contribution and credit within teams. Using a novel dataset linking Nobel Prizes to their definitive papers (22), we confirm the widening gap between those who contributed to groundbreaking research and those who are ultimately awarded the Nobel Prize. Over the past century (1905-2016), the average team size for Nobel Prize-winning papers increased three times, rising from 1.5 to 4.5. In contrast, the number of Nobel laureates within the team has remained relatively constant, averaging approximately 1.1. This growing disparity indicates an increase in the fraction of unrecognized team members, rising from 36% to 89% per prize-winning paper. To explore this further, we developed and analyzed the world's first large-scale observational dataset on author contributions, derived from LaTeX source files of 1.6 million papers authored by 2 million scientists. Using a validated credit allocation algorithm (10), we quantified individual credit shares for authors, enabling us to examine the relationship between contributions and credit while accounting for variables such as team size, career stage, and historical times. Our findings confirm that teams amplify the Matthew Effect—the gap between contribution and recognition widens in team-based science, disproportionately affecting emerging researchers collaborating with established colleagues, especially in large teams (18).

Results

**Many Contributed, but Few Were Recognized in Nobel Prize Teams.**

We analyzed 683 Nobel Prize-winning papers, representing the groundbreaking research that earned 432 laureates their prizes between 1905 and 2016. The connection between these research papers and Nobel laureates was established by analyzing the laureates' award speeches and identifying key references mentioned (22). This novel dataset enables us to examine how many individuals contributed to Nobel Prize-winning work, who they were, and whether they were ultimately recognized.

Our analysis reveals a significant growth in team size for Nobel Prize-winning work over the past century, with the decade-based average increasing threefold, from 1.5 in the early 1900s to 4.5 by 2016. However, despite this growth, the number of Nobel laureates per paper has remained steady at around 1.1 (Fig. 1). This indicates that while more contributors are essential for producing groundbreaking work at the forefront of research recognized by the Nobel Prize, a smaller proportion of team members receive recognition. The fraction of unrecognized team members has risen sharply, from 27% in the early years to 76% in modern times. This growing disparity underscores the widening gap between contribution and recognition, even within the most celebrated research teams worldwide.

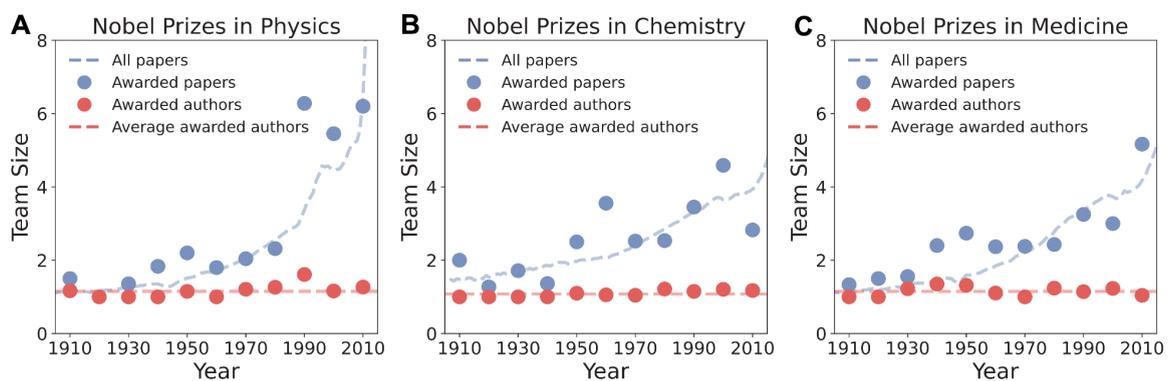

Figure 1. Many Contributed, but Few Were Recognized in Nobel Prize Teams. We analyzed 683 Nobel Prize-winning papers, representing the groundbreaking research that earned 432 laureates their prizes (1905-2016). This analysis draws on mentioning references from their award speeches (22). Blue dots show the average number of authors per decade for Physics (Panel A, 205 papers), Chemistry (Panel B, 214 papers), and Medicine (Panel C, 264 papers). This number increased threefold, rising from 1.5 to 4.5 across all three fields (from 1.3 to 6.2 for Physics, 2.0 to 2.8 for Chemistry, and 1.3 to 5.2 for Medicine). Consequently, the fraction of unrecognized team members grew from 27% in the early years (calculated as (1.5-1.1)/1.5) to 76% in modern times ((4.5-1.1)/4.5). Dashed blue lines indicate overall trends in team size estimated from nearly 36 million

papers in the Microsoft Academic Graph dataset (3.3 million in Physics, 10.5 million in Chemistry, and 21.7 million in Medicine). Red dots represent the average number of authors per decade on these groundbreaking research papers that were ultimately awarded Nobel Prizes, with dashed red lines indicating an overall average of approximately 1.1 across all three fields (1.2 for Physics, 1.1 for Chemistry, and 1.1 for Medicine).

**Large Teams Overshadow Individual Recognition**

With the recognition gap in Nobel Prizes in mind, we next turn to examine the core hypothesis of this paper: teams tend to divide and obscure credit, making it increasingly difficult to recognize individual contributions. Indeed, if scientific awards tend to recognize only one researcher within a team, it becomes unlikely that the primary contributor will be properly identified, particularly as team sizes grow. This challenge is further compounded by social biases, such as the Matthew Effect, where established scholars within the team are more likely to be recognized and remembered.

To test this hypothesis, we developed the world's first large-scale observational dataset on author contributions, derived from LaTeX source files of 1.6 million papers authored by 2 million scientists on arXiv.org. Established in the 1990s, arXiv.org is the largest preprint repository for STEM fields, including mathematics, statistics, computer science, and more, where LaTeX—a typesetting system widely used for formatting equations—is commonly adopted. Using this dataset, we identified author contributions by analyzing the paper LaTeX source code: if a paper contains a specific macro previously used by an author, we consider that author a potential contributor. A macro can be attributed to multiple authors if it matches their individual records. We calculated the number of unique macros contributed by each author and normalized it across all authors on a paper to estimate contribution share (Fig. 2A). We identified the contributions of 583,817 scientists across 730,914 papers (1991-2023). We validated this dataset against 469 self-reported author contributions collected from four journals, *Science*, *Nature*, *PNAS*, and *PLoS ONE*, and confirmed its high precision (0.87) and recall (0.71) in identifying paper-writing contributions. We collected demographic information on these scientists, including career age and gender. Career age was measured in years since their first publication, and gender was inferred from their first names using GPT-3.5-Turbo, which has been reported to achieve up to 96% accuracy in gender inference (23).

Next, we quantified individual credits within teams using the algorithm developed by Shen and Barabási (10). The algorithm estimates credit allocation among coauthors by analyzing citation patterns. For each focal paper, the algorithm identifies all co-cited papers and starts with an equal share of $1/n$ credit for each of the $n$ coauthors. It then iteratively updates each coauthor's credit by incorporating their relevant credit from co-cited papers as $1/m$ (where $m$ is the total number of authors on the co-cited paper), weighted by the co-citation strength between the co-cited paper and the focal paper. This approach captures a coauthor's visibility within a research field by considering the relevance and impact of their entire body of work, inferring their "socially perceived" author credits within the focal paper. In validation, the algorithm successfully identified laureates as the top recognized authors in 81% of multi-author Nobel Prize-winning papers (10). We applied this algorithm to the 2021 Microsoft Academic Graph archive, now part of OpenAlex, to calculate credit shares for 57 million scientists across 55 million journal articles, each with at least one citation, making them part of co-citation networks. These papers were then mapped to our LaTex macro dataset, resulting in a dataset of 121,492 scientists and 181,289 coauthored papers (1991-2021) with detailed information on contribution shares, credit shares, and author demographics. The matched dataset is much smaller than the LaTex macro dataset because approximately half of arXiv papers and a quarter of MAG journal papers lack DOIs, which we relied on for matching.

Analysis of this novel dataset reveals that teamwork's benefits and costs are distributed unevenly among team members based on author rank. Contribution is highest for the first author and declines with author rank, while credit increases and peaks for the last author. Larger teams exacerbate this disparity: in two-member teams, the first author does 0.2% more work and receives 8% less credit than the last author, but in six-member teams, they do 42% more work and receive 13% less credit (Fig. 2B-C). Combined, these effects heavily impact the first author—they contribute more but are less likely to be recognized, and this disparity grows with increasing team size.

What determines author rank? Further analysis reveals its correlation with career age (Pearson correlation coefficient: 0.27, $P < 0.001$, averaged across team sizes). This suggests that the observed inequality in recognition across author ranks is not merely about team roles but also reflects age dynamics in the scientific community, where older scientists often hold dominant positions such as teachers, mentors, and reviewers, making them more likely to be recognized than their younger colleagues (24, 25). Building on this insight, we use logistic regression models to examine how career age influences recognition probability—the likelihood of being identified as the "main brain," or the author deserving the most credit from coauthored papers. This approach mimics the process of selecting a Nobel Prize laureate based on team contributions. In the regression model, we account for contributions by identifying whether the author is the primary contributor, while controlling for confounding variables and their interactions.

Our results reveal that career age has the strongest positive effect in predicting recognition probability, ironically even exceeding the influence of being the primary contributor (Fig. 2D). Among other variables, being the last author also increases recognition, likely because it signals a dominant team role. Having a male name and working on more impactful papers both provide a statistically significant but slight advantage. Team size has the largest negative impact, as it dilutes individual credit, supporting our hypothesis and aligning with the empirical findings on credit distribution (Fig. 2B-C). Publication year has a negative effect, indicating that recognition in teamwork becomes harder over time, even when controlling for team size. This may be related to the rapid growth of papers, reducing citation and recognition opportunities (26).

We find that the interaction between career age and team size has a significant positive impact on recognition probability. This supports our observation that large teams disadvantage first authors and junior scientists. To further demonstrate this interactive effect, we created two logistic regression models to predict recognition probability and primary contributor probability separately, including all previously mentioned variables. The results show that while contributions remain relatively stable as career age increases, credit rises significantly, especially in larger teams (Fig. 2E-F).

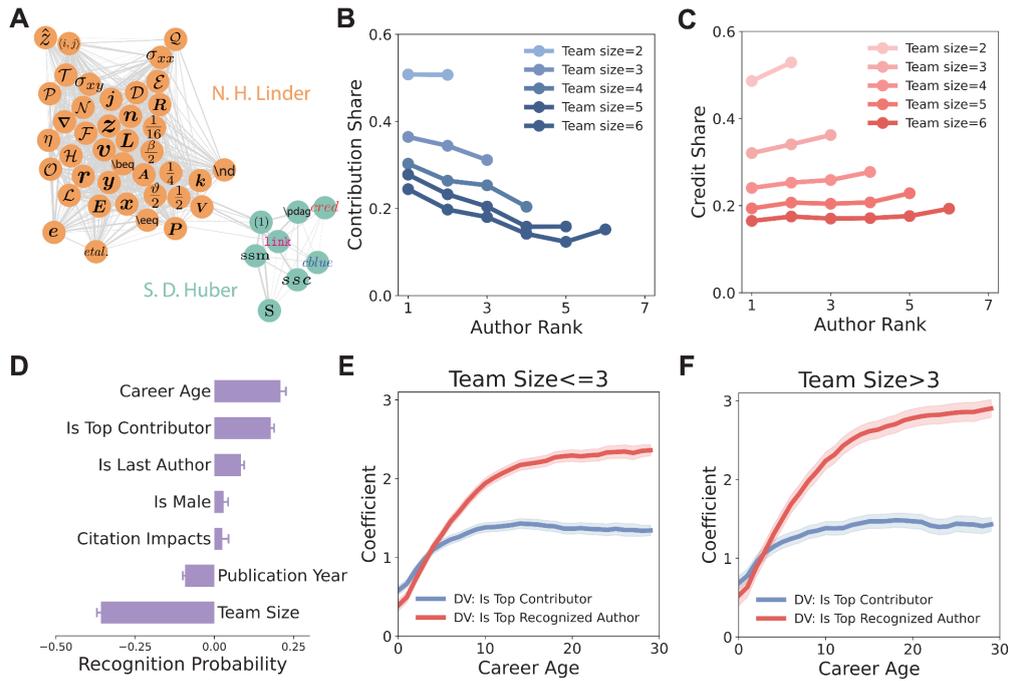

**Figure 2. Large Teams Overshadow Individual Recognition.** We analyzed and presented results from two datasets: the LaTeX Macro Dataset (Panel A) and the Author Contribution and Credit Dataset (Panel B-F). The first includes author-specific macros from 583,817 scientists across LaTeX source code underlying 730,914 arXiv papers (1991-2023). The second includes contribution shares, credit shares, and demographics for 121,492 scientists and 181,289 coauthored arXiv papers (1991-2021). Panel A shows a network of macros used in a collaborative paper by Huber and Lindner published in 2011. Nodes represent macros (green for Huber, yellow for Lindner), and links indicate co-usage with a paper across their publication records. Huber authored 42 papers with 26 unique macros, while Lindner authored 48 papers with 195 unique macros. In the focal paper, their contributions included 8 unique macros from Huber and 40 from Lindner, resulting in a contribution distribution of 1/6 and 5/6, respectively. While most links are distributed within individual scientists' clusters, some connect different clusters, suggesting learning from collaboration. For example, the link between the yellow and green clusters was created by Lindner after 2011, indicating that Lindner adopted new macros from Huber. Analysis of the Author Contribution and Credit Dataset reveals that contribution share decreases with author rank (B), while credit share increases (C). To conduct further analysis, we built three logistic regression models. The first predicts recognition probability—the likelihood of an author receiving the most credit for coauthored papers. This model includes four author-related variables (Career Age, CareerAge², Is Primary Contributor, Is Last Author, Is Male), four paper-related variables (logarithmic Citation Impacts, Publication Year, Team Size, and 15 discipline dummies), and 5 interaction terms (between Team Size and author-related variables), estimated using 468,346 author-paper pairs. The regression results are shown in (D), displaying the effect sizes of each variable within their typical ranges: 0–1 for binary variables, 0 to the median for CareerAge (11) and CitationImpacts (16), and the full observed ranges for Publication Year (1991–2021) and Team Size (2–7). The second and third models predict recognition probability and primary contributor probability, respectively, using the same dataset and variables. Results show that contributions remain stable with career age, but credit rises significantly, especially in larger teams (Fig. 2E-F).

## Discussion

This study builds on Merton's foundational observations on the contribution-credit paradox in science, extending them to the context of team-based science. Our findings illuminate an essential tension in modern science: while teams are essential for tackling complex scientific challenges, they also obscure individual credits and distort the reward system. This distortion particularly affects junior scientists in large teams.

Our findings have two key implications for junior scientists. First, they should be cautious about working in large and hierarchical teams, as such environments may not only suppress individual autonomy and creativity (13, 14, 27), but also divide and obscure credit. Second, they should focus on building a cohesive body of work that gains recognition, rather than spreading their efforts across diverse topics, which can hinder recognition (28, 29).

Our work has broader implications for the scientific community, including academic journal editors, research institution leaders, and policymakers. Addressing the challenge of recognition in teamwork requires systemic changes in how scientific contributions are evaluated and rewarded (30). While mandating contribution statements is a promising step, it is insufficient to counteract biases. This limitation originates from the fundamental conflict between the scalability of scientific contributions and the scarcity of the community's collective attention and memory. Simply documenting who does what does in the team does not ensure these contributions will be remembered. Further research is needed in this area, and, perhaps more importantly, recognition practices—such as individual-based scientific awards—must evolve to mitigate the costs of teamwork.

This study is grounded in a novel, verified dataset of author contributions derived from LaTeX source codes. By extending beyond journals that require contribution statements, this dataset provides unique insights into the dynamics of teamwork in fields heavily represented on arXiv, such as Physics, Mathematics, and Computer Science. However, this narrow disciplinary focus limits the generalizability of our findings. Future research should expand these methods to include other repositories, such as bioRxiv, to capture a broader spectrum of scientific practices and team structures.

**Material and Methods**

**Nobel Prize Dataset.** This dataset comprises 683 research papers (1905–2016) referenced by 432 Nobel Prize laureates in their award speeches, developed by Li et al. (22). It includes 205 papers in Physics (1905–2016), 214 in Chemistry (1908–2016), and 264 in Medicine (1912–2016). These papers were identified from the 2021 archived version of the Microsoft Academic Graph (https://zenodo.org/records/6511057), from which basic information such as author count, publication year, and field of study labels was extracted.

**LaTeX Macro Dataset.** This dataset contains LaTeX source files for 1,600,627 preprints authored by 2,012,092 unique scientists (1991–2023). Author contributions are identified by analyzing LaTeX source codes, following the approach outlined by Pei et al. (manuscript in preparation): if a paper includes a macro previously used by an author, they are considered a potential contributor. Macros can be attributed to multiple authors if matched to their records. Unique macro counts were calculated for each author and normalized across all authors on a paper to estimate contribution shares. This process identified contributions for 583,817 scientists across 730,914 papers (1991–2023). Validation against 469 self-reported author contributions from four journals, *Science*, *Nature*, *PNAS*, and *PLoS ONE*, confirmed high precision (0.87) and recall (0.71) for identifying paper-writing contributions.

**Author Contribution and Credit Dataset.** We applied the credit allocation algorithm developed by Shen and Barabási (10) to infer author credits. Validation showed the algorithm successfully identified laureates as the top recognized authors in 81% of multi-author Nobel Prize-winning papers (10). We then applied the algorithm to the 2021 Microsoft Academic Graph archive, which contains 90,054,610 journal papers and 91,947,512 unique authors. To ensure credit could be inferred, we selected 54,764,230 papers with at least one citation, making their authors part of co-citation networks. These papers were mapped to our arXiv dataset, yielding a final dataset of 121,492 scientists and 181,289 coauthored papers

(1991–2021) with detailed information on contribution shares, credit shares, and author demographics. Additionally, we collected demographic information of these authors, including career age (years since first publication) and gender (inferred from names using the Python package *gender-guesser*).

**Acknowledgments**

We are grateful for support from the National Science Foundation grant SOS:DCI 2239418 (L.W.) and SES-2152437, the National Institutes of Health R01GM148816 and 1P20GM152280 and from the Alfred P. Sloan Foundation G-2024-22494 (D.G.).

**Competing Interests**

The authors declare to have no competing interests.

**References**

1.  R. K. Merton, The Matthew effect in science. The reward and communication systems of science are considered. *Science* **159**, 56–63 (1968).

2.  S. Wuchty, B. F. Jones, B. Uzzi, The increasing dominance of teams in production of knowledge. *Science* **316**, 1036–1039 (2007).

3.  D. J. de S. Price, *Little science, big science* (Columbia University Press New York, 1963).

4.  E. Leahey, From Sole Investigator to Team Scientist: Trends in the Practice and Study of Research Collaboration. (2016). https://doi.org/10.1146/annurev-soc-081715-074219.

5.  B. F. Jones, The Rise of Research Teams: Benefits and Costs in Economics. *J. Econ. Perspect.* **35**, 191–216 (2021).

6.  B. K. Ryu, The demise of single-authored publications in computer science: A citation network analysis. *arXiv [cs.DL]* (2020).

7.  N. T. Hagen, Harmonic publication and citation counting: sharing authorship credit equitably - not equally, geometrically or arithmetically. *Scientometrics* **84**, 785–793 (2010).

8.  G. Sivertsen, R. Rousseau, L. Zhang, Measuring scientific contributions with modified fractional counting. *J. Informetr.* **13**, 679–694 (2019).

9.  L. Waltman, N. J. van Eck, Field-normalized citation impact indicators and the choice of an appropriate counting method. *J. Informetr.* **9**, 872–894 (2015).

10. H. W. Shen, A. L. Barabási, Collective credit allocation in science. *the National Academy of Sciences* (2014).

11. P. Donner, A validation of coauthorship credit models with empirical data from the contributions of PhD candidates. *Quant. Sci. Stud.* 1–14 (2020).

12. L. Egghe, R. Rousseau, G. Van Hooydonk, Methods for accrediting publications to

authors or countries: Consequences for evaluation studies. *J. Am. Soc. Inf. Sci.* **51**, 145–157 (2000).

13. F. Xu, L. Wu, J. Evans, Flat Teams Drive Scientific Innovation. *Proc. Natl. Acad. Sci. U. S. A.* **119** (2022).

14. Y. Lin, C. B. Frey, L. Wu, Remote collaboration fuses fewer breakthrough ideas. *Nature* **623**, 987–991 (2023).

15. C. Haeussler, H. Sauermann, Division of labor in collaborative knowledge production: The role of team size and interdisciplinarity. *Res. Policy* **49**, 103987 (2020).

16. V. Larivière, *et al.*, Contributorship and division of labor in knowledge production. *Soc. Stud. Sci.* **46**, 417–435 (2016).

17. H. Sauermann, C. Haeussler, Authorship and contribution disclosures. *Sci. Adv.* **3**, e1700404 (2017).

18. M. Andalón, C. de Fontenay, D. K. Ginther, K. Lim, The rise of teamwork and career prospects in academic science. *Nat. Biotechnol.* **42**, 1314–1319 (2024).

19. S. Milojević, F. Radicchi, J. P. Walsh, Changing demographics of scientific careers: The rise of the temporary workforce. *Proc. Natl. Acad. Sci. U. S. A.* **115**, 12616–12623 (2018).

20. S. Chatrchyan, *et al.*, Observation of a new boson at a mass of 125 GeV with the CMS experiment at the LHC. *Phys. Lett. B* **716**, 30–61 (2012).

21. B. P. Abbott, *et al.*, Observation of Gravitational Waves from a Binary Black Hole Merger. *Phys. Rev. Lett.* **116**, 061102 (2016).

22. J. Li, Y. Yin, S. Fortunato, D. Wang, A dataset of publication records for Nobel laureates. *Sci. Data* **6**, 33 (2019).

23. A. Domínguez-Díaz, M. Goyanes, L. de-Marcos, V. P. Prado-Sánchez, Comparative analysis of automatic gender detection from names: evaluating the stability and performance of ChatGPT versus Namsor, and Gender-API. *PeerJ Comput. Sci.* **10**, e2378 (2024).

24. R. K. Merton, *The Sociology of Science: Theoretical and Empirical Investigations* (University of Chicago Press, 1973).

25. P. Azoulay, J. S. Graff Zivin, J. Wang, Superstar Extinction. *Q. J. Econ.* **125**, 549–589 (2010).

26. J. S. G. Chu, J. A. Evans, Slowed canonical progress in large fields of science. *Proc. Natl. Acad. Sci. U. S. A.* **118** (2021).

27. L. Wu, D. Wang, J. A. Evans, Large teams develop and small teams disrupt science and technology. *Nature* **566**, 378–382 (2019).

28. E. Berkes, M. Marion, S. Milojević, B. A. Weinberg, Slow convergence: Career

impediments to interdisciplinary biomedical research. *Proc. Natl. Acad. Sci. U. S. A.* **121**, e2402646121 (2024).

29. E. Leahey, C. M. Beckman, T. L. Stanko, Prominent but less productive: The impact of interdisciplinarity on scientists' research. *Adm. Sci. Q.* **62**, 105–139 (2017).

30. D. Hicks, P. Wouters, L. Waltman, S. de Rijcke, I. Rafols, Bibliometrics: The Leiden Manifesto for research metrics. *Nature* **520**, 429–431 (2015).